\documentclass[twocolumn]{aastex61}
\usepackage{amsmath,amstext}
\usepackage[T1]{fontenc}
\usepackage{apjfonts} 
\usepackage[figure,figure*]{hypcap}

\shorttitle{Baryonic effects in cosmic shear tomography}
\shortauthors{I. Mohammed and N. Gnedin}

\begin{document}

\title{Baryonic effects in cosmic shear tomography: PCA parametrization and importance of extreme baryonic models}

\author{Irshad Mohammed}\thanks{mohammed@fnal.gov}
\affiliation{Theoretical Astrophysics Group, Fermi National Accelerator Laboratory, Batavia, IL 60510, USA}
\affiliation{Kavli Institute for Cosmological Physics, The University of Chicago, Chicago, IL 60637 USA}
\author{Nickolay Y.\ Gnedin}\thanks{gnedin@fnal.gov}
\affiliation{Theoretical Astrophysics Group, Fermi National Accelerator Laboratory, Batavia, IL 60510, USA}
\affiliation{Kavli Institute for Cosmological Physics, The University of Chicago, Chicago, IL 60637 USA}
\affiliation{Department of Astronomy \& Astrophysics, The University of Chicago, Chicago, IL 60637 USA}

\begin{abstract}
Baryonic effects are amongst the most severe systematics to the tomographic analysis of weak lensing data which is the principal probe in many future generations of cosmological surveys like LSST, Euclid etc.. Modeling or parameterizing these effects is essential in order to extract valuable constraints on cosmological parameters. In a recent paper, Eifler et al. (2015) suggested a reduction technique for baryonic effects by conducting a principal component analysis (PCA) and removing the largest baryonic eigenmodes from the data.  In this article, we conducted the investigation further and addressed two critical aspects. Firstly, we performed the analysis by separating the simulations into training and test sets, computing a minimal set of principle components from the training set and examining the fits on the test set. We found that using only four parameters, corresponding to the four largest eigenmodes of the training set, the test sets can be fitted thoroughly with an RMS $\sim 0.0011$. Secondly, we explored the significance of outliers, the most exotic/extreme baryonic scenarios, in this method. We found that excluding the outliers from the training set results in a relatively bad fit and degraded the RMS by nearly a factor of 3. Therefore, for a direct employment of this method to the tomographic analysis of the weak lensing data, the principle components should be derived from a training set that comprises adequately exotic but reasonable models such that the reality is included inside the parameter domain sampled by the training set. The baryonic effects can be parameterized as the coefficients of these principle components and should be marginalized over the cosmological parameter space.

\end{abstract}

%%%%%%%%%%%%%%%%%%%%%%%%%%%%%%%%%%%%%%%%%%%%%%%%%%

%%%%%%%%%%%%%%%%% BODY OF PAPER %%%%%%%%%%%%%%%%%%

\section{Introduction}\label{sec:intro}

There are diverse pieces of evidence from existing observations, such as Cosmic Microwave Background CMB \citep{2003ApJS..148..175S, 2014A&A...571A..16P}, Baryon Acoustic Oscillations BAO \citep{2010MNRAS.401.2148P, 2012MNRAS.427.3435A, 2013MNRAS.433.1202S}, Gravitational Lensing \citep{1993ApJ...404..441K} etc., indicating that the major component of the matter content of the Universe, over 80$\%$, is the so-called dark matter, which interacts only gravitationally. The mostly collisionless nature of dark matter has empowered our computational capabilities to accurately simulate the Universe on linear and quasi-linear scales using dark matter only (DMO) simulations. On the other hand, doing so has introduced a major systematic uncertainty in several observational probes due to the different dynamics of baryonic matter, a small but non-negligible contribution to the total matter budget, at sufficiently small scales. These small scales contain a significant amount of information on the cosmological parameters; to extract that information one has, therefore, to correctly model these systematics. Without such modeling,  constraints on cosmological parameters and interpretation of observational data may be biased \citep{2013PhRvD..87d3509Z}. 

Bending of light due to intervening matter (a lens) is referred to as gravitational lensing (for a thorough review see \cite{1999ARA&A..37..127M, 2001PhR...340..291B, 2006glsw.conf..269S}). If the gravitational potential of the lens is large, and the geometry of the observer-lens-source is favorable, multiple images of the source can be observed, the phenomenon also known as strong gravitational lensing (SL). On the other hand, if the potential of the lens is not strong enough, and/or the geometry is not perfect, only the shape of the sources get distorted. This phenomenon is known as weak gravitational lensing (WL). The distortions in the shape of the background galaxies are referred to as {\it shear}. The shear signal is small compared to the noise and the intrinsic shape/ellipticity of the source, nearly 1$\%$. Therefore, it can only be measured statistically, by averaging out the noise and the intrinsic ellipticity of a sample of sources together, assuming there is no preferred ellipticity or orientations of the background galaxies. The most commonly used statistic is the two-point correlation function, or its Fourier transform, the power spectrum. The tomographic analysis of the shear power spectra in different redshift bins is amongst the most promising tool to constrain cosmological parameters, including the equation of state of the dark energy \citep{wme13}.

As lensing is sensitive to the total matter content of the lens and does not differentiate between dark matter and baryons, this probe is fairly unbiased to any baryonic effects observationally. However, theoretical modeling of the shear power spectra relies on modeling the distribution of all matter in the Universe, which is strongly affected by the effects of baryonic physics at small scales. There have been many attempts to model these baryonic effects using hydrodynamical simulations \citep{2006ApJ...640L.119J, 2008ApJ...672...19R, 2010MNRAS.405..525G, 2011MNRAS.415.3649V, 2011MNRAS.417.2020S, 2013MNRAS.434..148S} and theoretical modeling \citep{2004APh....22..211W, 2004ApJ...616L..75Z, 2014MNRAS.445.3382M, 2014arXiv1410.6826M}.

In a recent paper, \cite{2015MNRAS.454.2451E} proposed a mitigation technique for baryonic effects by performing a principal component analysis (PCA) and removing the largest baryonic eigenmodes from the data. This technique is equivalent to de-baryonising the shear power spectra data, and then fitting it with dark matter only (DMO) models to constrain the cosmological parameter. In this article, we took the analysis further and discussed two important aspects. First, we divided the simulations into training and test sets, performing the PCA on training sets and analyzing the fits on both training and test sets. This shows the universality of the results. Secondly, we discussed the importance of outliers (or the most exotic/extreme baryonic scenarios) in this analysis. 

This paper is organized as follows: In section \ref{sec:theory} we review the theoretical ingredients of the tomographic analysis of the shear power spectrum. In section \ref{sec:sims} we briefly discuss different hydrodynamical simulations used in this work. In section \ref{sec:results} we briefly discuss the concept of dimensionality reduction and present the main results of this exercise. In section \ref{sec:analysis} we describe the full analysis pipeline and in section \ref{sec:outliers} we emphasize the importance of the outliers. Finally, we discuss our results in section \ref{sec:discussion}.

%=============================================================================
\section{Theoretical Review - Modeling Shear Power Spectrum}\label{sec:theory}

The distortion of the source shape due to WL can be described by two quantities: shear $\gamma$ and convergence $\kappa$. The convergence $\kappa$ is the local isotropic part of the deformation matrix and can be expressed as:

\begin{equation}
	\kappa(\vec{\theta}) = \dfrac{1}{2} \vec{\bigtriangledown} \cdot \vec{\alpha}(\vec{\theta}),
\end{equation}
where $\vec{\theta}$ is the angular coordinate on the sky and $\vec{\alpha}$ is the deflection angle.
If we know the redshift of the source galaxies, extra information can be gained by separating the sources in various redshift bins. This method is referred to as lensing {\it tomography} and is very useful to gain extra constraints on cosmology from the evolution of the weak lensing power spectra \citep{1999ApJ...522L..21H, 2004MNRAS.348..897T, 2009MNRAS.395.2065T}. In cosmological setting, the convergence field can be represented as the weighted projection of the mass distribution integrated along the line of sight in the $i$th redshift bin,

\begin{equation}
	\kappa_i(\vec{\theta}) = \int_0^{\chi_H} g_i(\chi) \delta(\chi \vec{\theta}, \chi) d\chi,
\end{equation}
where $\delta$ is the total 3-dimensional matter overdensity, $\chi$ is the comoving distance, and $\chi_H$ is the comoving distance to the horizon. The lensing weights $g_i(\chi)$ in the $i$-th redshift bin with comoving distance range between $\chi_i$ and $\chi_{i+1}$ are given by:

\begin{equation}
g_i(\chi) = \begin{cases} \dfrac{g_0}{\bar{n}_i} \dfrac{\chi}{a(\chi)} \int_{{\rm max}(\chi_i, \chi)}^{\chi_{i+1}} n_s(\chi^{\prime})\dfrac{dz}{d\chi^{\prime}} \dfrac{(\chi^{\prime} - \chi) }{\chi^{\prime}} d\chi^{\prime}, & \chi \le \chi_{i+1} \\
			0, & \chi > \chi_{i+1} \end{cases}
\end{equation}
with $a(\chi)$ being the scale factor at comoving distance $\chi$. Also,
\begin{equation}
	g_0 = \dfrac{3}{2} \Omega_m H_0^2
\end{equation}
and
\begin{equation}
	\bar{n}_i = \int_{\chi_i}^{\chi_{i+1}} n_s(\chi(z)) \dfrac{dz}{d\chi^{\prime}} d\chi^{\prime}.
\end{equation}
Here $n_s(\chi(z))$ is the distribution of sources in redshift. We assume a source distribution along the line of sight of the form:
\begin{equation}
	n_s(z) = n_0 \times 4z^2 \exp\left(-\dfrac{z}{z_0}   \right)
\end{equation}
with $n_0 = 1.18 \times 10^{9} $ per unit steradian and $z_0$ is fixed such that the mean redshift of the source distribution is $z_m=3z_0$ and corresponding projected source density $n_g$ resembles the experiment (for a more detailed review, see \cite{2009MNRAS.395.2065T}),
\begin{equation}
	\int_0^{\infty} n_s(z)dz = \bar{n}_g.
\end{equation}
Table \ref{tbl:1} shows the weak lensing survey parameters for a stage-III and stage-IV survey in this setting. Finally the shear power spectrum between redshift bins $i$ and $j$ can be computed as:
\begin{equation}
	C_{ij}(\ell) =  \int_0^{\chi_H} \dfrac{g_i(\chi) g_j(\chi)}{ \chi^2} P\left(\dfrac{\ell}{\chi},\chi \right)d \chi,
\end{equation}
where $P$ is the 3D matter power spectrum . Larger $\ell$ corresponds to the smaller scale and the large contribution of $C_{\ell}$ at higher $\ell$ comes from non-linear clustering.

\begin{table}
\centering
\begin{tabular}{| l| l c c c|}
\hline
 Survey & Stage & $z_0$ & $z_m$ & $\bar{n}_g$\\
\hline
 DES & III & 0.23 & 0.7 & 51 \\  
 LSST & IV & 0.4 & 1.2 & 100\\
 \hline
\end{tabular}
\caption{Weak lensing survey parameters for a stage-III (DES) and stage-IV (LSST) like survey.}
 \label{tbl:1}
\end{table}

%=============================================================================
\section{Baryonic Effects in Simulations}\label{sec:sims}

In this section, we describe various hydrodynamical simulations used to study different baryonic corrections on the tomographic analysis of the weak lensing shear power spectra for a DES and LSST like survey.  We worked with the same sets of simulations examined by \cite{2015MNRAS.454.2451E}. We thank Tim Eifler for kindly providing us with the shear power spectra corresponding to various DMO and hydrodynamical simulations for DES and LSST like surveys. All spectra are computed for five tomographic bins, there are a total of $5(5+1)/2=15$ spectra available (five auto-spectra and ten cross-spectra). Each of these spectra are computed for 12 multi-pole bins ($\ell$) equally spaced in logarithm. Therefore, the length of each of the simulation vector is $15\times 12=180$.

The full simulation set is comprised of three subsets: 
\begin{description}
\item[\bf OWLS simulations:] This suite of simulations contributes total nine different baryonic scenarios, which differ for their cooling, supernovae and AGN feedback. For a detailed prescription of these simulations and the OWLS project, see \cite{2010MNRAS.402.1536S, 2011MNRAS.415.3649V, 2013PhRvD..87d3509Z, 2013MNRAS.434..148S}. 
\item[\bf ART08 simulations:] This subset of simulations performed with the Adaptive Refinement Tree (ART) code \cite{k99,kkh02,2008ApJ...672...19R} contributes total two different baryonic scenarios, one of which is treated in the non-radiative regime and does not allow any star or galaxy formation, whereas the other simulations allow these processes. For a detailed description of these simulations see \cite{2008ApJ...672...19R}. These simulations were called "Rudd simulations" in \cite{2015MNRAS.454.2451E}.
\item[\bf ART14 simulations:] Three more baryonic scenarios modeled with the ART code are contributed by this subset, the first set contains adiabatic hydrodynamic processes, the second set radiative cooling, but not heating, with primordial abundance of hydrogen and helium, whereas the third set contains extreme radiative cooling with cooling function similar to solar-metallicity gas. These simulations were called "Gnedin simulations" in \cite{2015MNRAS.454.2451E}.
\end{description}

\begin{figure*}
    \centering
    \includegraphics[width=0.8\textwidth]{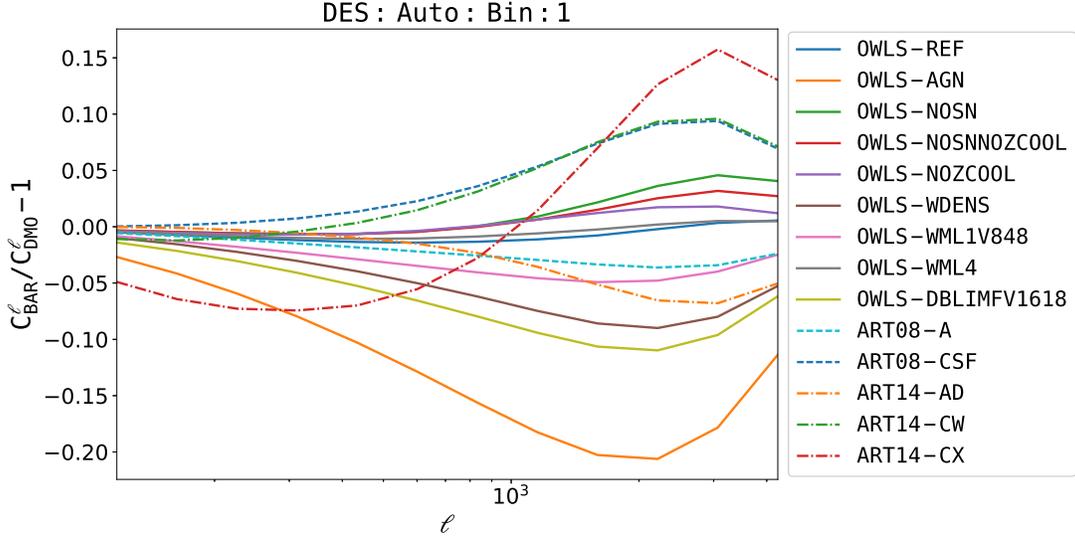}
    \caption{Shear spectra in the first redshift bin for DES like survey. Solid lines: OWLS simulations, dashed lines: ART08 simulations, and dashed-dotted lines: ART14 simulations. In section \ref{sec:outliers} we describe $\mathtt{ART14-CX}$ simulation as the outlier, or the most exotic one as one can see in this figure (dashed-dotted red line). }
\label{fig:orig:des}
\end{figure*}

Figure \ref{fig:orig:des} shows the deviation of the shear power spectra of each of these baryonic scenarios from the dark matter only case for the auto-spectra in the first redshift bin for a stage - III (DES-like) survey. For a stage - IV (LSST-like) survey, we have shear spectra for all but ART08 simulations. So there are total 12 scenarios available for a LSST-like survey and 14 for a DES-like survey.

%=============================================================================
\section{Results}\label{sec:results}

\subsection{Dimensionality reduction}

The advancement in our observational and computational abilities has led us into the era of big datasets, which are observed and/or simulated everyday. These datasets not only have an enormous sample size but also have high dimension, i.e., a large number of variables measured/computed in each observation/simulation. These variables are often correlated, in a linear or a non-linear way, which gives us the ability to look for fewer derived variables that can be used to represent the full original dataset with lower dimension. This method is referred to as {\it Dimensionality Reduction}. In this text, we use the term {\it variables} interchangeably with {\it features} or {\it attributes}.

Mathematically, a given dataset with $p$ variables, $\mathbf{x}=(x_1, x_2, x_3, ..., x_p)$ can be reduced to a new dataset with $k$ variables, $\mathbf{s}=(s_1, s_2, s_3, ..., s_k)$, where $k<p$, with the ability to reconstruct the original dataset using some criterion. The techniques are broadly classified into two categories: linear and non-linear, depending on whether the mathematical form of the mapping from $\mathbf{x}$ to $\mathbf{s}$ (or vice-versa) is a linear or non-linear function, respectively. One very popular linear dimensional reduction technique is the standard Principal Component Analysis (PCA). For a more general overview of both linear and non-linear technique see \cite{FODOR2002}.

PCA is amongst the most commonly used linear dimensionality reduction technique. It attempts to reduce the dimension of the data by finding an alternate basis for the original variables with the largest variance-preserving the covariance matrix of the variables. The basis is in the form of orthogonal linear combinations, also referred to as Principal Components (or PCs), normalized to the square-root of the corresponding eigenvalue.

\subsection{Analysis}\label{sec:analysis}

We start by defining baryons contrast ($\delta C_{ij}(\ell)$) (also interchangeably referred to as {\it boost}) for the weak lensing auto and cross shear power spectra in different redshift bins, which represent the relative deviation from the DMO case,

\begin{equation}
    \delta C_{ij}(\ell) = \dfrac{C^{\rm BAR}_{ij}(\ell)}{C^{\rm DMO}_{ij}(\ell)} - 1.
\end{equation}

\begin{figure}
    \centering
    \includegraphics[width=0.46\textwidth]{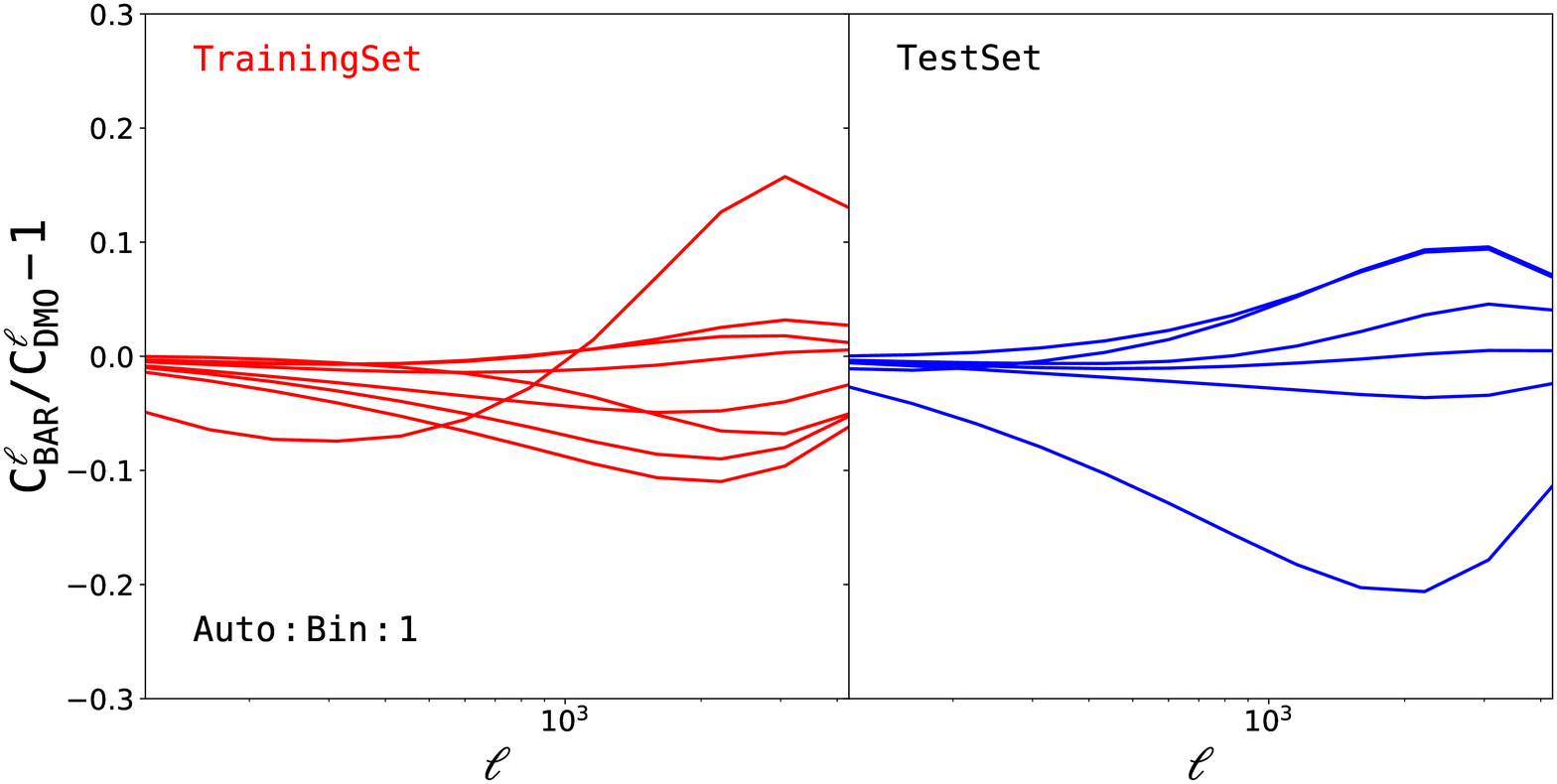}
    \includegraphics[width=0.46\textwidth]{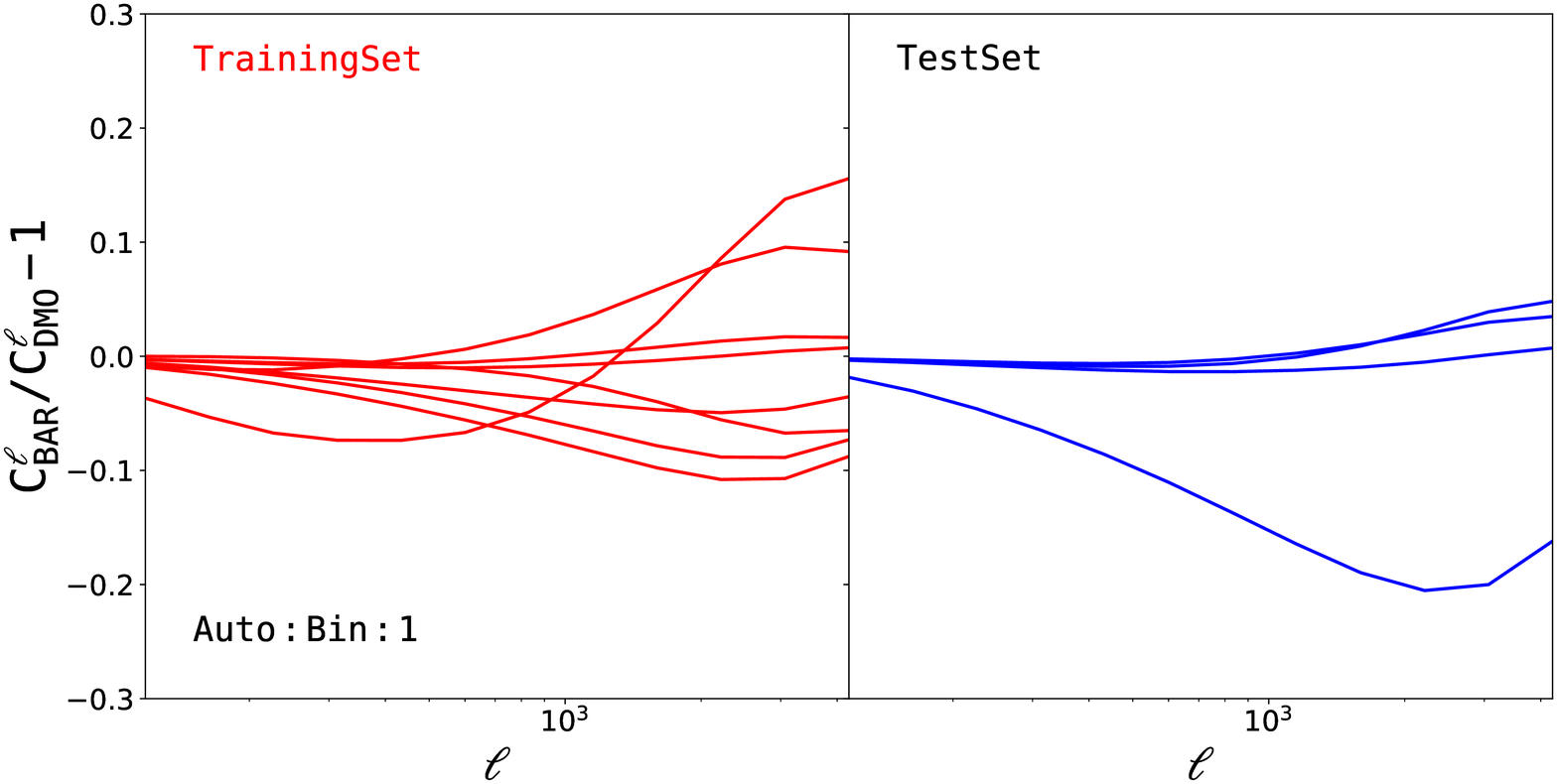}
    \caption{Training and test sets for DES (upper-row) and LSST (bottom-row) like surveys.}
    \label{fig:spectra}
\end{figure}

Second, we split the simulation sets of each case, DES and LSST, into {\it training} and {\it test} sets. The training set includes the subset of all baryonic scenarios used to compute the minimal set of principal components PCs. For each case, we use a subset of eight baryonic scenarios as the training set, and the remaining four simulations for LSST case (or six for DES case) as the test set. The first and second row of Figure \ref{fig:spectra} shows the baryon contrast for both training (left panel) and test (right panel) sets for DES and LSST cases respectively.

\begin{figure}
	\centering
	\includegraphics[width=0.46\textwidth]{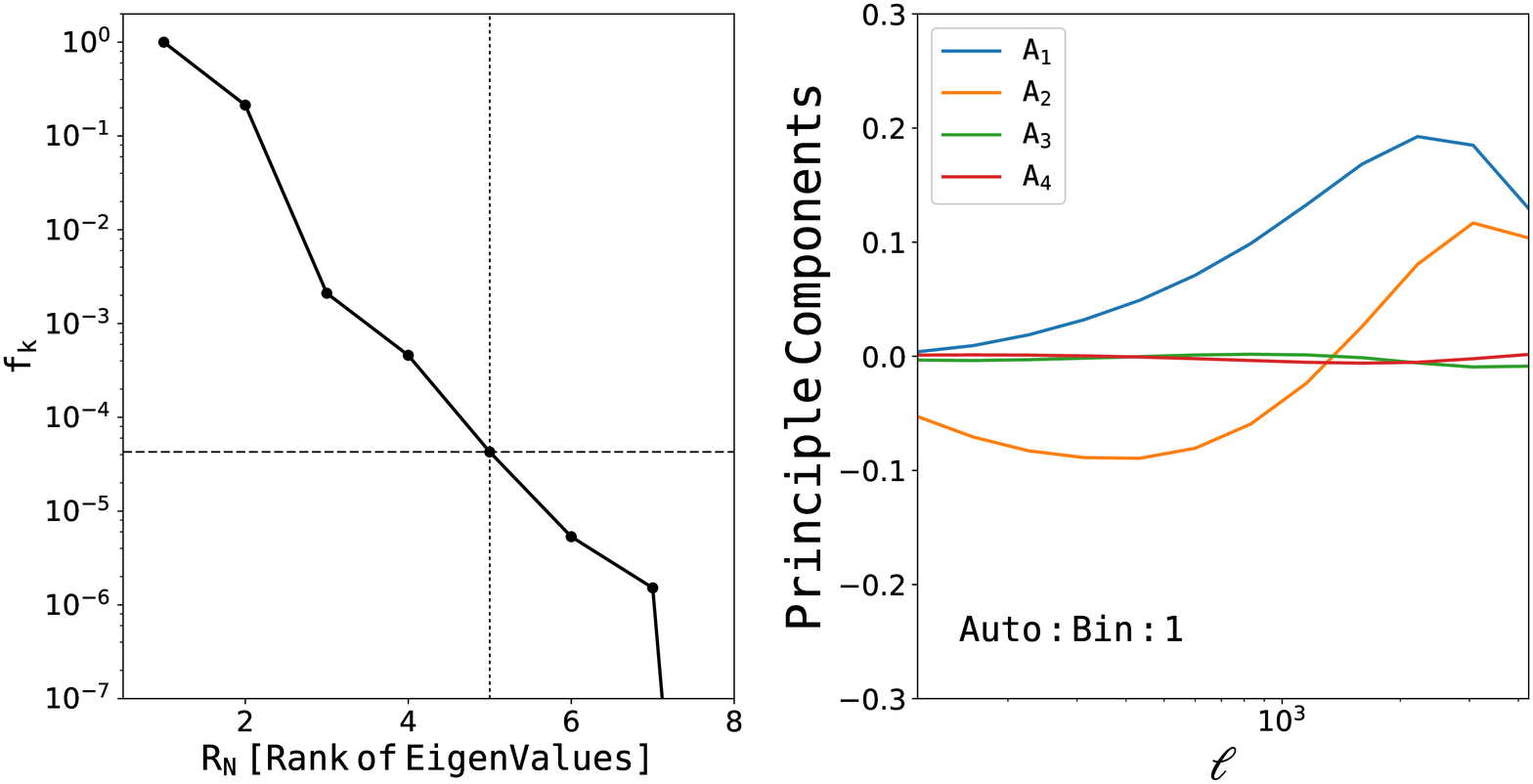}
	\includegraphics[width=0.46\textwidth]{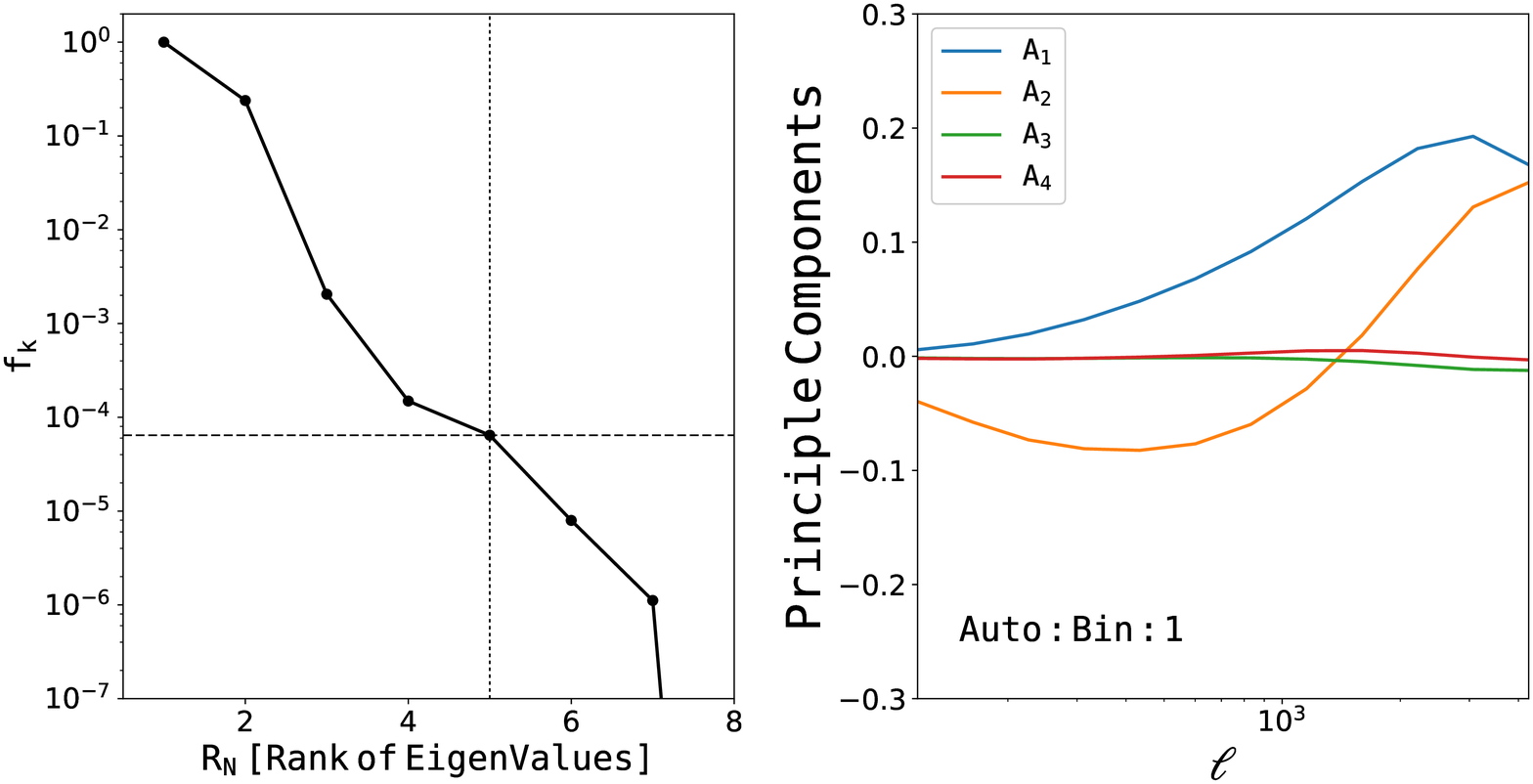}
    \caption{Eigenvalue fraction (left panels) and the four principle components (right panel) for DES (Upper-row) and LSST (Bottom-row) like surveys. The PCs are shown only for the first tomographic bin.}
	\label{fig:ev}
\end{figure}

Next, we compute the $180\times 180$ covariance matrix of the training sample and determine the ordered eigenvalues (total 180, however only first 8 are non-vanishing as there are only 8 samples in the training set). The eigenvalue fraction is defined as the fraction of the total covariance matrix trace \emph{not} accounted for  by first $k$ eigenvalues,

\begin{equation}
	f_k = 1-\frac{\sum_{i=1}^k E_{i}}{{\rm Tr}(\delta C_{ij})},
    \label{eq:fk}
\end{equation}
where $E_{i}$ is the $i$th eigenvalue. The eigenvalue fraction quantifies the fractional residue left over after one removes the first $k$ eigenmodes.  The left panels of Figure \ref{fig:ev} shows the eigenvalue fraction for DES (upper row) and LSST (bottom row) like surveys. The first eigenvalue contains nearly 80$\%$ of the total fraction of the variance, therefore after removing the first eigenmode, the eigenvalue fraction is $\sim 0.2$. Similarly, if we remove the second eigenvalue, the eigenvalue fraction drops to $\sim 0.002$. For both cases, we chose $k=4$ as the new reduced dimension, i.e.\ we remove the first four eigenmodes, which gives the eigenvalue fraction of $\sim 0.0001$. We will discuss this particular choice below. The right panels of figure \ref{fig:ev} show the four PCs extracted from the respective training sets normalized to the square root of the respective eigenvalues. 
% * <gnedin@fnal.gov> 2017-06-28T16:42:28.062Z:
% 
% How are principal components normalized? 
% 
% ^ <creativeishu@gmail.com> 2017-06-29T15:50:41.987Z:
% 
% The default normalization of the PC is to the square root of the corresponding eigenvalue. I mentioned this in the last line of section 4.1
%
% ^ <creativeishu@gmail.com> 2017-06-29T17:49:46.638Z.

\begin{figure}
	\centering
	\includegraphics[width=0.46\textwidth]{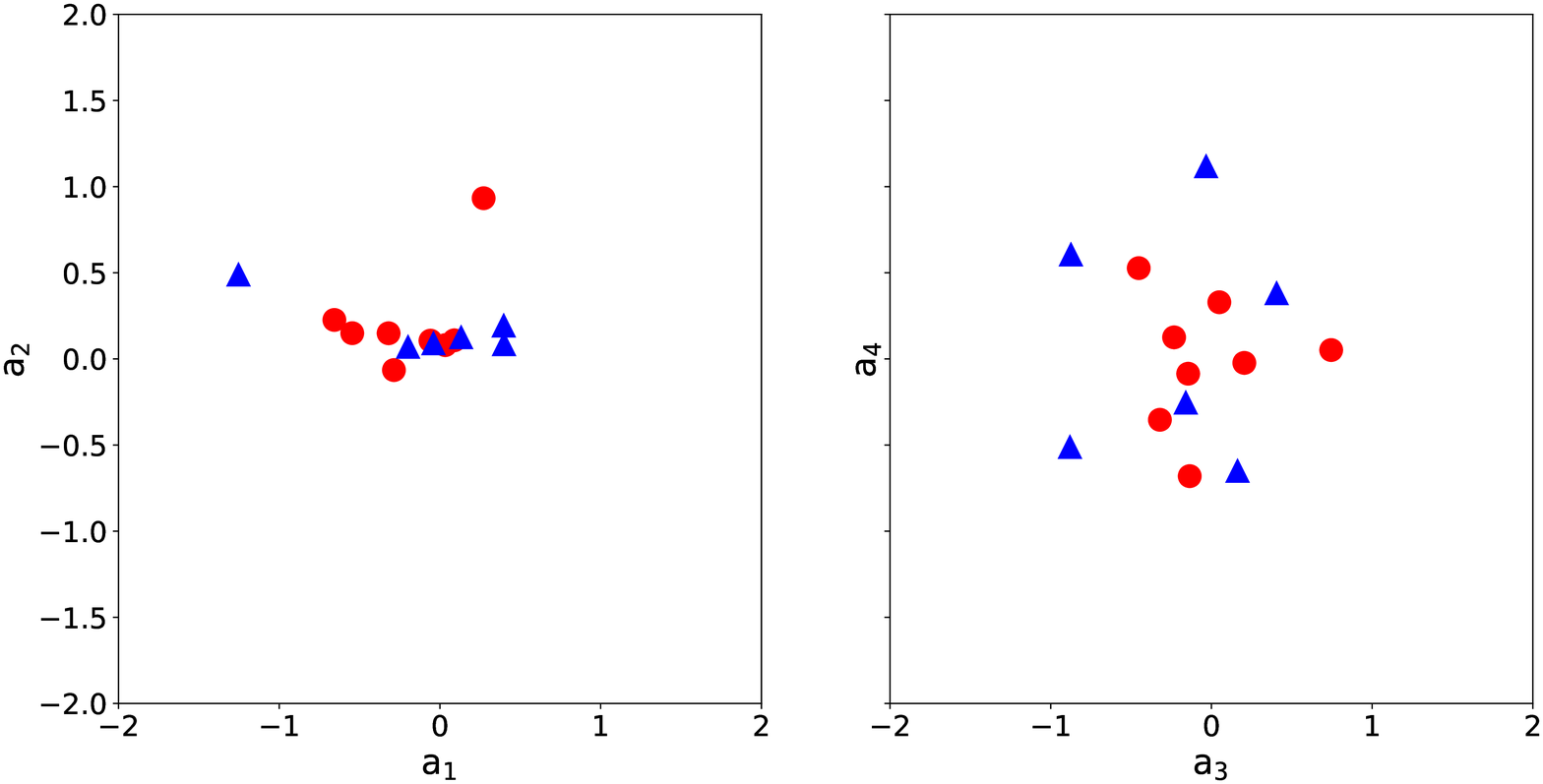}
	\includegraphics[width=0.46\textwidth]{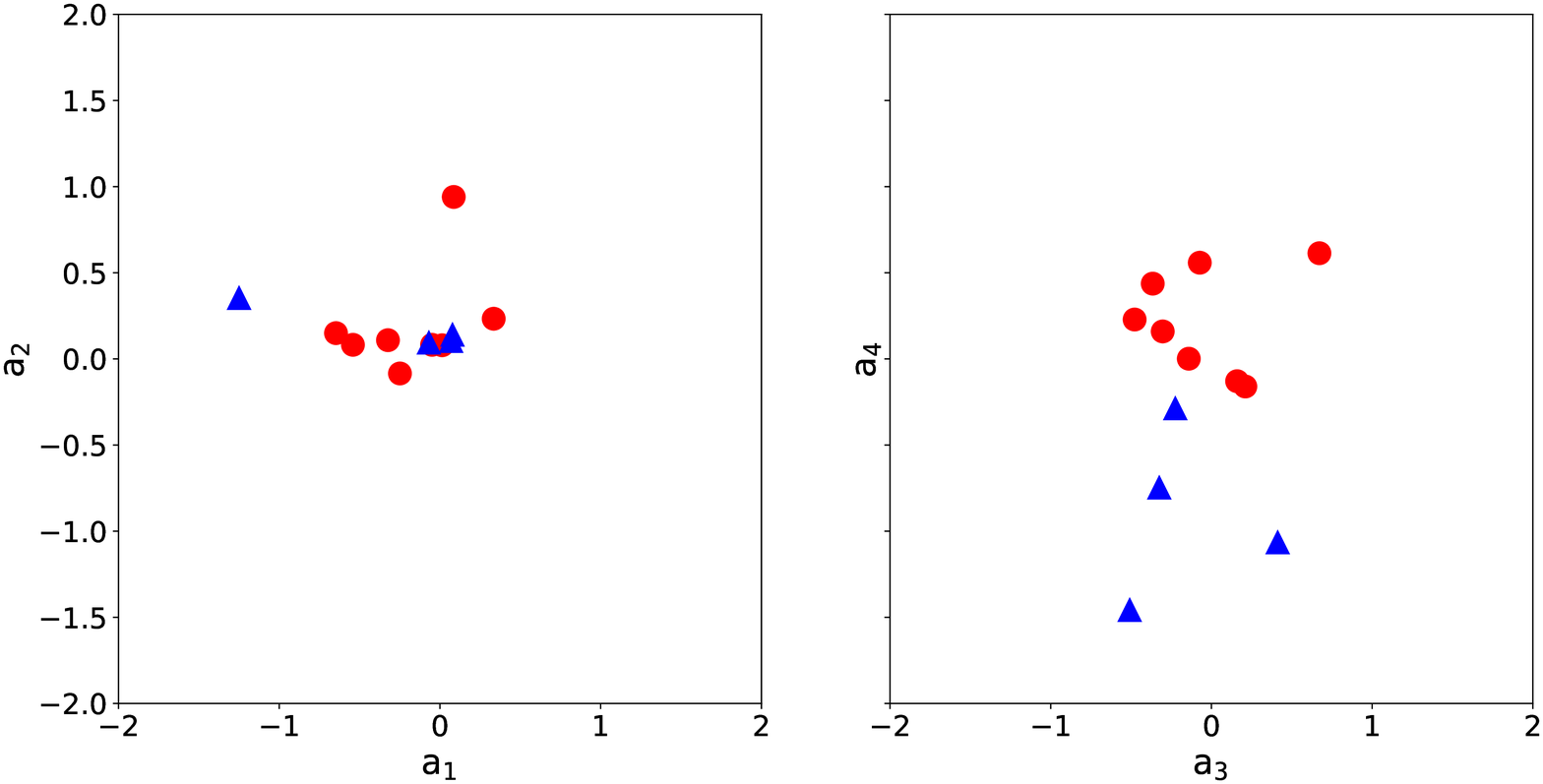}
    \caption{Best fit coefficients for the training (red circles) and test (blue triangles) for DES (upper row) and LSST (bottom row) like survey.}
	\label{fig:coeff}
\end{figure}

\begin{figure}
	\centering
	\includegraphics[width=0.46\textwidth]{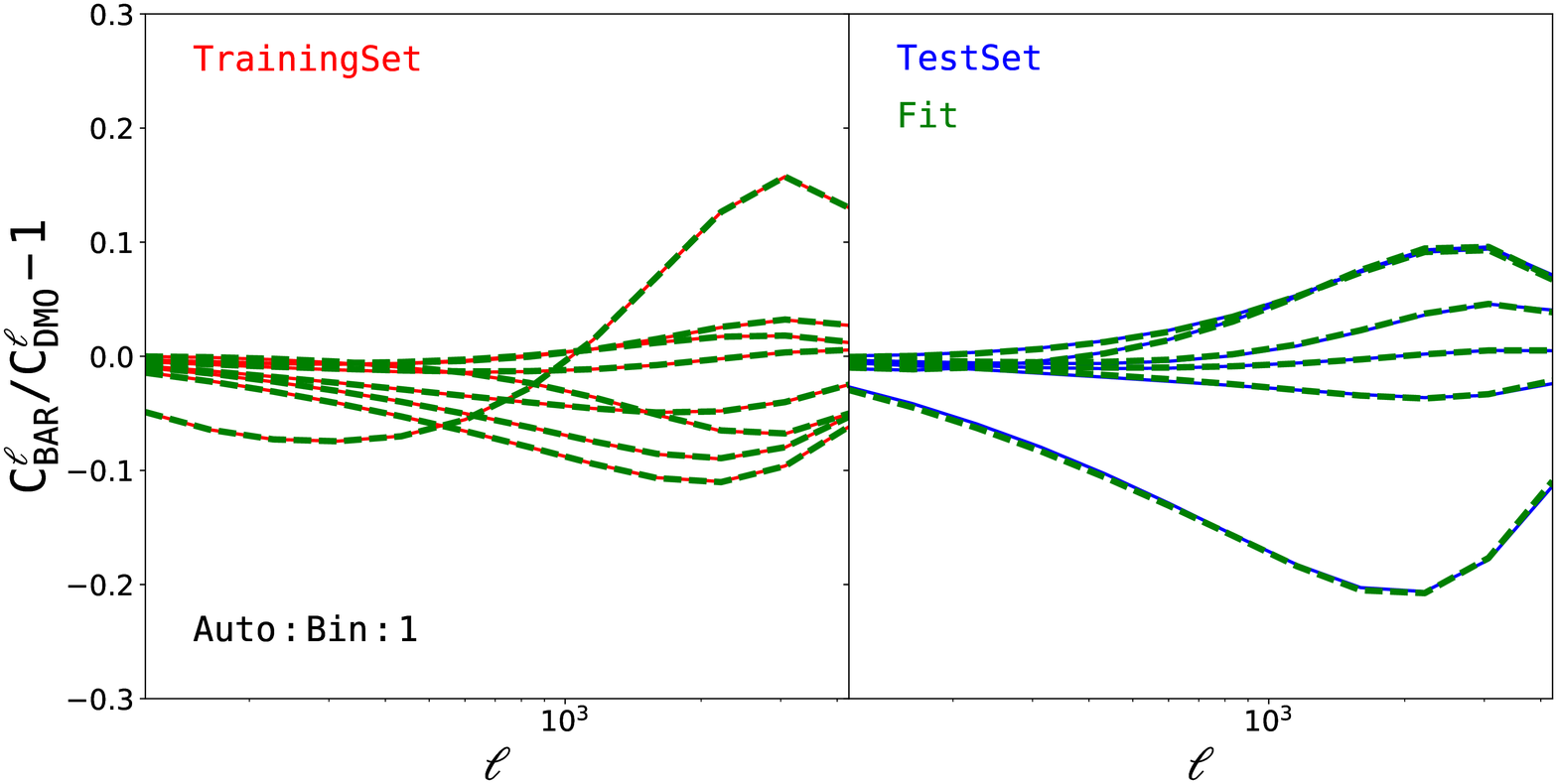}
	\includegraphics[width=0.46\textwidth]{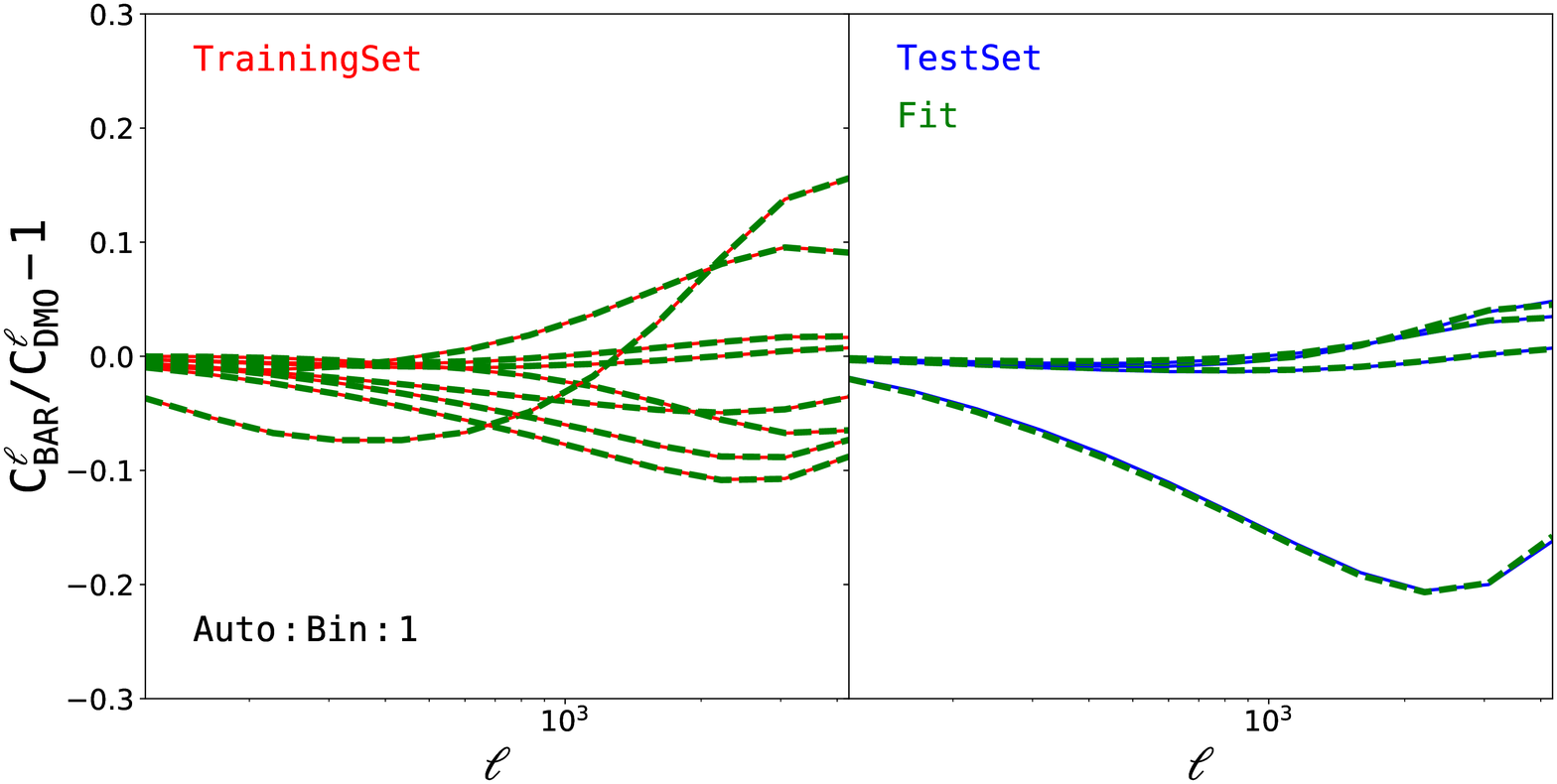}
    \caption{Fits to baryon contrasts (in dashed-green lines) corresponding resulting from summing the first four eigenmodes (with coefficients shown in Figure \ref{fig:coeff}) for DES (upper row) and LSST (bottom row) like surveys.}
	\label{fig:fits}
\end{figure}

Using only first four eigenmodes,  we fit both the training and the test sets with a linear combination of the PCs, 
\begin{equation}
    \delta C_{ij}^{\rm Fits}(\ell) = \sum_{i=1}^{4} a_i A_i,
\end{equation}
where, $A_i$ are the four PCs (or the four largest eigenvectors) as  shown in figure \ref{fig:ev}, and $a_i$ are the coefficients. Figure \ref{fig:coeff} shows the best fit coefficients for the DES and LSST cases and figure \ref{fig:fits} shows the corresponding fits. 

Finally,  the full model of the shear power spectra, including parametric form of the baryonic effects, can be presented as

\begin{equation}
    C^{\rm BAR}_{ij}(\ell) = C^{\rm DMO}_{ij}(\ell)\ (1+\delta C_{ij}^{\rm Fits}(\ell)).
\end{equation}
This model contains a total of four free parameters, in addition to the cosmological parameters.

%---------------------------------------------------------------------------
\begin{figure}
	\centering
	\includegraphics[width=0.46\textwidth]{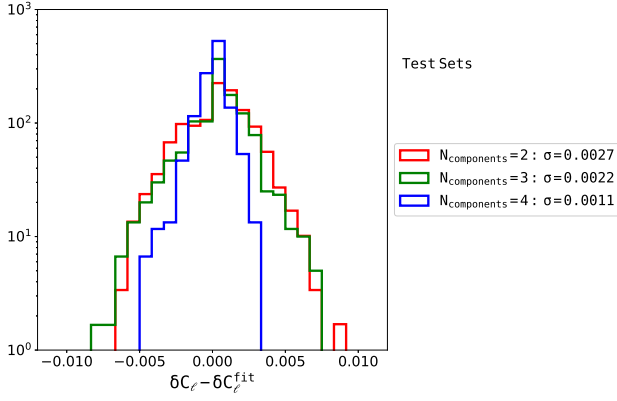}
    \caption{Distributions of rms errors for 2, 3, and 4-PC fits to the actual simulated baryonic contrasts for the fiducial case. There is significant gain in using 4 components over 2 or 3.}
	\label{fig:errs}
\end{figure}

The particular choice of four eigenvalues that we adopt above is dictated by the precision requirements for an LSST-like survey. In Figure \ref{fig:errs} we show the distributions of fitting errors (differences between dashed and solid lines in Fig.\ \ref{fig:fits}) for 2, 3, and 4 principal components ($k=2$, 3, and 4 in Eq.\ \ref{eq:fk}). Only with four principal components, the errors go safely below 1\%. In addition, there is nearly a factor of 2 gain in increasing number of components from 3 to 4. We, therefore, use 4 components as a fiducial setup.

\subsection{Importance of outlier models}\label{sec:outliers}

Outliers of the dataset refer to the most extreme baryonic scenarios, which, never-the-less, cannot be excluded on purely physical grounds. In our simulation sets, we mark $\mathtt{ART14-CX}$ simulation as such an outlier, which is also evident by its shape (see figure \ref{fig:orig:des}). In our analysis in the previous section we included this outlier in the training set, and therefore, the PCs contain its signatures. In this section, we explore its importance in the case where the outlier is not present in the training set but exists in the test set. We perform this exercise for LSST case only, i.e. with eight training examples and four test examples.

\begin{figure}
    \centering
    \includegraphics[width=0.46\textwidth]{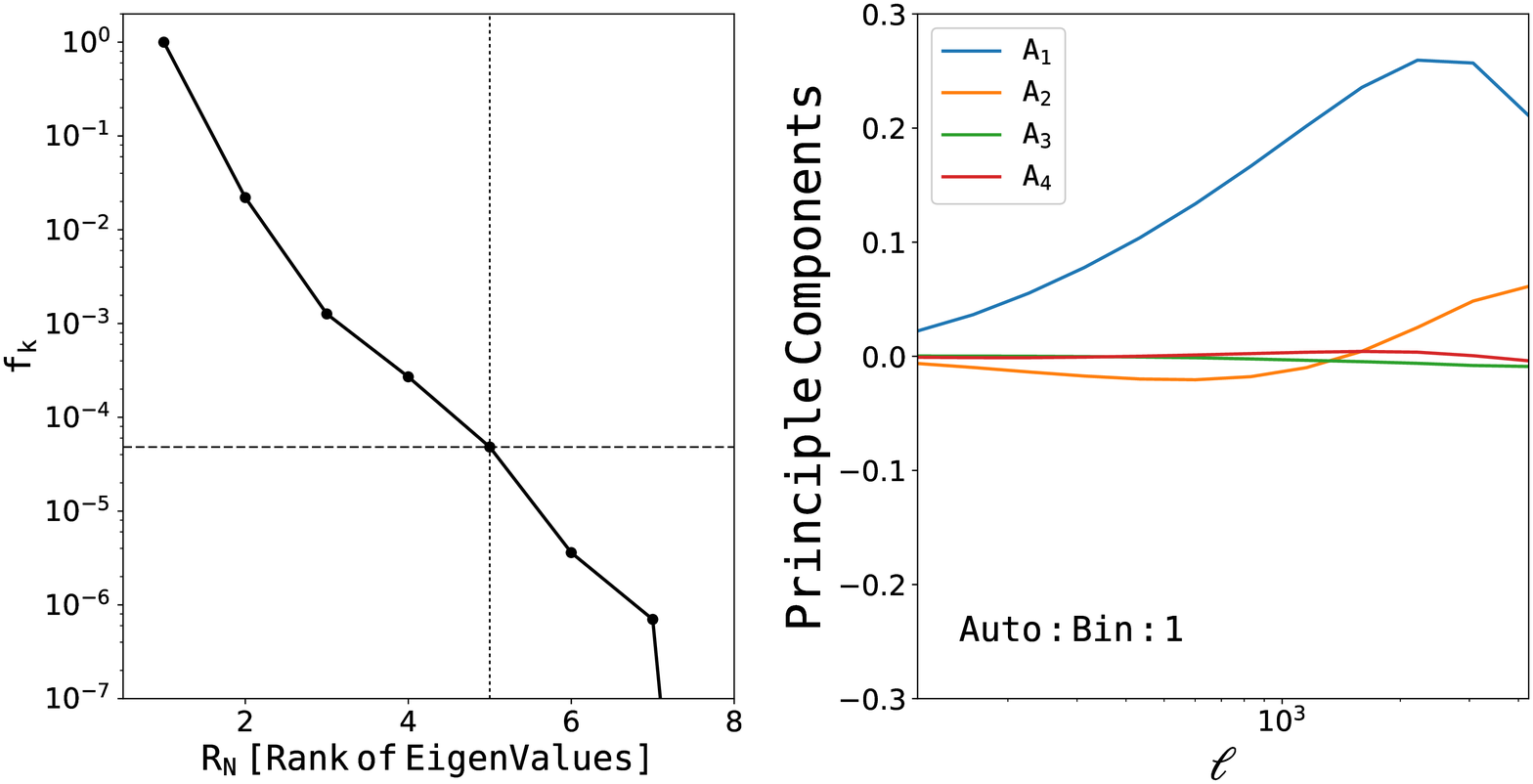}\\
    \includegraphics[width=0.46\textwidth]{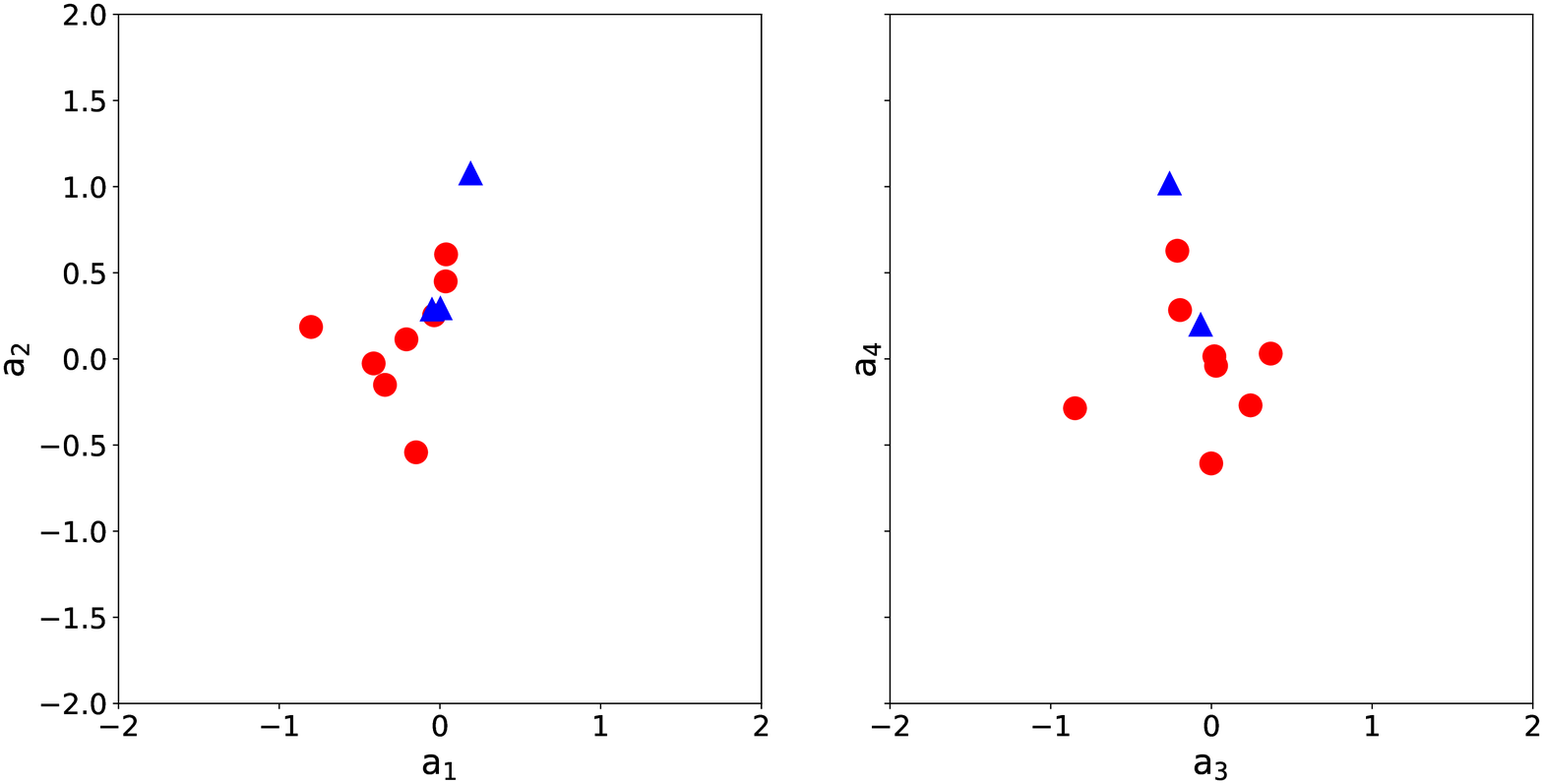}
    \caption{Analysis of the LSST case with the outlier model located in the test set and not in the training set. Top row: the eigenvalue fraction and the PCs. Bottom row: the corresponding best fit coefficients. The blue triangle in the top area ( at coordinates 0,3) is the outlier model.}
    \label{fig:lsst:outlier}
\end{figure}

\begin{figure}
    \centering
    \includegraphics[width=0.46\textwidth]{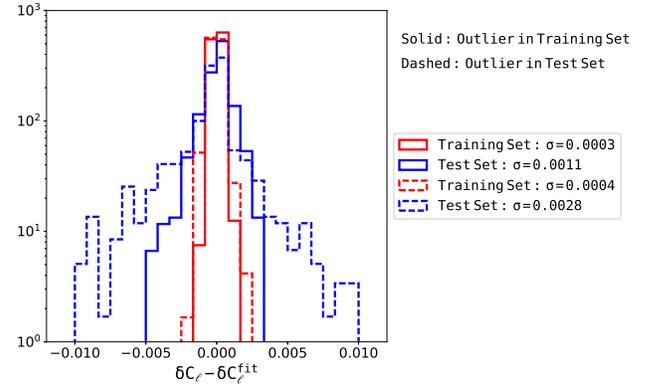}
    \caption{Distributions of rms errors for four-PC fits to the actual simulated baryonic contrasts for the case with the outlier model in the training set (solid lines) and for the case of the outlier model in the test set (dashed lines).  In the latter case errors increase by a factor of 3. }
    \label{fig:lsst:hist}
\end{figure}

Figure \ref{fig:lsst:outlier} shows the the eigenvalue fraction, PCs, and best-fit coefficients for this case.  In order to make a quantitative comparison with the previous case, we also show in Figure \ref{fig:lsst:hist} the precision with which approximations with four principal components are able to fit the actual simulated baryonic contrasts. When the outlier model is moved from the training set into the test set, errors in the fitted baryonic contrasts increase about threefold, both in the RMS sense and maximum errors.

%=============================================================================
\section{Summary and discussion}\label{sec:discussion}

The \cite{2015MNRAS.454.2451E} approach to mitigating baryonic effects in weak lensing is based on (1) building up a training set of simulations that model baryonic effects, (2) computing several (3-4) principal components of baryonic contrast, and (3) using these components as fitting functions with free parameters to marginalize over the baryonic modifications in the observed power spectra. The success or failure of such an approach obviously depends on the fidelity of the training set.

In this paper, we showed that such a training set has to be sufficiently broad in a sense of including extreme, but still physical meaningful models. It is not just enough to include simulations that attempt to model baryonic contrast as accurately as possible, because in that case the principal components may not be sensitive enough to other "degrees of baryonic freedom", i.e. variations in the power spectra that are not faithfully captured by the simulations (either due to numerical limitations or missing some physical ingredients). However, if "outlier" models are included, the precision of \cite{2015MNRAS.454.2451E} approach increases multi-fold. 

In the specific example that we consider here the outlier model (ART14-CX) adopts the maximal realistic value for the gas cooling function (that of the solar metallicity gas). While this is certainly extreme, such a model cannot be excluded from pure physical reasons: cooling is overestimated in the low-density IGM, but the weak lensing signal is dominated by clustered, higher density structures, where such strong cooling is not an obvious overestimate.

Hence, the ideal training set should include sufficiently exotic, but still not outright unreasonable models, to offer sufficiently justified assurance that the reality is included inside the parameter region sampled by the training set. Building such a training set is an important and immediate goal for the cosmological weak lensing community.

%%%%%%%%%%%%%%%%%%%% Acknowledgements %%%%%%%%%%%%%%%%%%

\section*{Acknowledgements}

This manuscript has been authored by Fermi Research Alliance, LLC under Contract No. DE-AC02-07CH11359 with the U.S. Department of Energy, Office of Science, Office of High Energy Physics. The United States Government retains and the publisher, by accepting the article for publication, acknowledges that the United States Government retains a non-exclusive, paid-up, irrevocable, world-wide license to publish or reproduce the published form of this manuscript, or allow others to do so, for United States Government purposes.

\bibliographystyle{yahapj}
\def\apj{ApJ}
\def\fcp{fcp}
\def\apjl{ApJL}
\def\aj{AJ}
\def\mnras{MNRAS}
\def\aap{A\&A}
\def\nat{Nature}
\def\pasj{PASJ}
\def\prd{PRD}
\def\physrep{Physics Reports}
\def\jcap{JCAP}
\bibliography{ms.bib}

% Don't change these lines
% \bsp	% typesetting comment
\label{lastpage}
\end{document}